# TRAITEMENT DES DONNEES MANQUANTES AU MOYEN DE L'ALGORITHME DE KOHONEN

Marie Cottrell, Smaïl Ibbou, Patrick Letrémy

SAMOS-MATISSE UMR 8595
90, rue de Tolbiac
75634 Paris Cedex 13

*Résumé :*
*Nous montrons comment il est possible d'utiliser l'algorithme d'auto organisation de Kohonen pour traiter des données avec valeurs manquantes et estimer ces dernières. Après un rappel méthodologique, nous illustrons notre propos à partir de trois applications à des données réelles.*

*Mots-clés :*
*Analyse de données, cartes de Kohonen, Données manquantes.*

*Summary :*
*We show how it is possible to use the Kohonen self-organizing algorithm to deal with data which contain missing values and to estimate them. After a methodological recall, we illustrate our purpose from three real databases applications.*

## 1. INTRODUCTION

Le traitement des données avec observations manquantes est un problème concret et toujours embarrassant lorsqu'il s'agit de données réelles. En effet dans les applications, on est très souvent en présence d'observations pour lesquelles on ne dispose pas de l'ensemble des valeurs des variables descriptives, et ceci se produit pour de nombreuses raisons : erreurs de saisie, rubriques non renseignées dans des enquêtes, valeurs aberrantes qu'on préfère supprimer, données recueillies difficilement, statistiques officielles non disponibles, etc.

La plupart des logiciels statistiques (comme SAS par exemple) suppriment purement et simplement les observations incomplètes, mais si cela n'a pas de conséquences pratiques lorsqu'on dispose de données très nombreuses, cela peut supprimer tout intérêt à l'étude si le nombre de données restantes est trop faible.

Pour éviter de supprimer ainsi les données, on peut remplacer une valeur manquante par la moyenne de la variable correspondante, mais cette moyenne peut être une très mauvaise approximation dans le cas où la variable présente une grande dispersion.

Il est alors dans ce cadre très intéressant de constater que l'algorithme de Kohonen (ainsi que l'algorithme de classification de Forgy) supporte parfaitement la présence de données manquantes, sans qu'il soit nécessaire de les estimer préalablement. Nous nous intéressons particulièrement ici à l'algorithme de Kohonen, puisque qu'il généralise l'algorithme stochastique des centres mobiles (version stochastique de l'algorithme de Forgy), et qu'en raison de ses propriétés de visualisation, il sera plus simple de « voir » les résultats.

On trouvera dans la thèse de Smaïl Ibbou (1998) un chapitre consacré à cette question, mais cela n'avait pas donné lieu à publication jusqu'à maintenant. C'est pour cette raison que nous avons pensé utile de présenter l'état de l'art sur cette question. Les exemples sont traités à l'aide du programme écrit par Patrick Letrémy en IML-SAS et disponible sur le site du SAMOS (http://samos.univ-paris1.fr).

## 2. ADAPTATION DE L'ALGORITHME DE KOHONEN A DES DONNEES INCOMPLETES.

On suppose que les observations sont des vecteurs à valeurs réelles de dimension $p$

Lorsqu'on présente un vecteur de données incomplet $x$, on détermine d'abord l'ensemble $M_x$ des numéros des composantes manquantes. $M_x$ est un sous ensemble de $\{1, 2, ..., p\}$. Si $(C_1, C_2, ..., C_n)$ est l'ensemble des vecteurs-codes à cet instant, on calcule le vecteur-code gagnant $C_{i_O(x)}$ associé à $x$ et sa classe, en posant

$$i_0(C, x) = Arg \min_i \|x - C_i\|$$

où la distance $\|x - C_i\|^2 = \sum_{k \notin M_x} (x_k - C_{i,k})^2$ est calculée sur les composantes présentes dans le vecteur $x$.

On peut utiliser les vecteurs avec données manquantes de deux façons.

a) Si l'on souhaite les utiliser au moment de la construction des vecteurs-codes, à chaque étape, une fois déterminé le numéro de l'unité gagnante, la mise à jour des vecteurs codes (le gagnant et ses voisins) ne porte que sur les composantes présentes dans le vecteur.

2) Si l'on dispose de suffisamment de données pour pouvoir se passer des vecteurs incomplets pour construire la carte, on peut aussi se contenter de classer, après construction de la carte, les vecteurs incomplets en les affectant dans la classe dont le vecteur code est le plus proche, au sens de la distance restreinte aux composantes présentes.

Cela donne d'excellents résultats, dans la mesure bien sûr où une variable n'est pas complètement absente ou presque, et aussi dans la mesure où les variables sont corrélées, ce qui est le cas dans la plupart des jeux de données réelles.

On pourra voir de nombreux exemples dans la thèse de Smaïl Ibbou (1998) et aussi dans Gaubert, Ibbou et Tutin (1996).

## 3. ESTIMATION DES VALEURS MANQUANTES

Quelle que soit la méthode utilisée pour utiliser les données avec valeurs manquantes, une des propriétés les plus intéressantes de l'algorithme, et qu'il est possible d'estimer a posteriori les valeurs manquantes.

Si $M_x$ est l'ensemble des numéros des composantes manquantes de l'observation $x$, et si $x$ est classé dans la classe $i$, pour chaque indice $k$ de $M_x$, on estime $x_k$ par :
$$\hat{x}_k = C_{i,k}.$$

Comme la fin de l'apprentissage de l'algorithme de Kohonen se fait à « zéro voisin », on sait que les vecteurs codes sont asymptotiquement proches des moyennes de leur classe. Cette méthode d'estimation consiste donc à estimer les valeurs manquantes d'une variable par sa moyenne dans la classe.

Il est clair que cette estimation est d'autant plus précise que les classes formées par l'algorithme sont homogènes et bien séparées les unes des autres. De nombreuses simulations ont montré tant dans le cas de données artificielles que de données réelles, qu'en présence de variables corrélées, la précision de ces estimations est remarquable. Voir à ce sujet le chapitre 5 de la thèse de Smaïl Ibbou (1998).

Pour augmenter la précision, il propose dans sa thèse de réaliser plusieurs versions de l'algorithme de Kohonen, et de prendre la moyenne des estimations obtenues dans chaque carte.

## 4. EXEMPLES PRESENTES

Dans la suite, nous présentons trois exemples tirés de données réelles.

Le premier exemple porte sur des données socio-économiques recueillies sur 182 pays en 1996. Les observations sont de trois sortes : 114 données complètes où chaque pays est décrit par 7 variables numériques, 52 pays avec une valeur manquante sur 7, et 16 pays où le nombre de valeurs manquantes est supérieur ou égal à 2. Les données correspondantes sont du domaine public et ont souvent été utilisées, à la suite du travail de François Blayo et Pierre Demartines(1991). Voir par exemple Cottrell, de Bodt, Verleysen (2002), Dans cet exemple, nous montrons comment utiliser les observations incomplètes de deux façons différentes, comme précisé au paragraphe 2.

Le second exemple porte sur des données immobilières de 205 communes d'Ile-de-France, en 1988, représentées par 15 variables. Dans cet exemple réel, issu d'une étude commandée par la DATAR et traitée par Christian Tutin, Patrice Gaubert et Smaïl Ibbou (1996), seules 5 communes n'ont pas de valeurs manquantes, 150 communes ont moins de 12 données manquantes, et 55 en ont 12 ou plus (sur 15). Il était donc essentiel de pouvoir utiliser les observations AVEC données manquantes.

Enfin, le troisième exemple est un exemple classique en analyse de données, tiré du Que-sais-je ? de Bouroche et Saporta , « L'analyse des données » ( 1980). Il s'agit de la structure de dépenses de l'Etat, mesurées sur 24 années entre 1872 et 1971, par un vecteur de dimension 11. Dans cet exemple, nous avons artificiellement supprimé des valeurs présentes dans les données originelles, d'une valeur sur 11 à 8 valeurs sur 11, choisies aléatoirement, pour évaluer la précision des estimations obtenues en remplaçant ces valeurs par les valeurs correspondantes des vecteurs codes associés, comme expliqué dans le paragraphe 3.

# 5. LES DONNEES SOCIO-ECONOMIQUES.

On a mesuré en 1996, 7 variables se rapportant à la situation macro-économique of 182 pays. Ce type de données a été utilisé pour la première fois dans le contexte des cartes de Kohonen par François Blayo et Pierre Demartines (1991).

Les variables mesurées sont :
La croissance annuelle de la population en % (ANCRX), le taux de mortalité infantile (en pour mille) (TXMORT), le taux d'illettrisme en % (TXANAL), l'indice de fréquentation scolaire au second degré (SCOL2), lePNB par habitant exprimé en dollars (PNBH), le taux de chômage en % (CHOMAG), le taux d'inflation en % (INFLAT).

Ces pays peuvent aussi être classés a priori d'après une classification classique en économie fondée sur l'indice de développement humain (NIVIDH), variable qualitative à 6 niveaux codés par des libellés : faible1 (1), faible2 (2), moyen1 (3), moyen2 (4), fort1 (5), fort2 (6).

Parmi l'ensemble des 182 pays considérés, seuls 114 ne présentent pas de données manquantes, 52 ont une valeur manquante, tandis que 16 ont plus de 2 valeurs manquantes.

On utilise donc les 114 + 52 = 166 pays complets ou presque pour construire la carte de Kohonen, et on classe ensuite les 16 pays restant. Les données sont centrées et réduites comme classiquement. La figure 1 montre la répartition des vecteurs codes après convergence de l'algorithme, la figure 2 montre le contenu des classes, en ne tenant compte que des 166 pays utilisés pour le calcul des vecteurs codes. On voit que les pays riches se trouvent dans le coin en haut et à gauche, les pays très pauvres en haut et à droite. Les pays à grande inflation sont en bas de la carte, à proximité des pays riches se trouvent des pays intermédiaires, comme ceux qui faisaient partie du camp socialiste, etc.

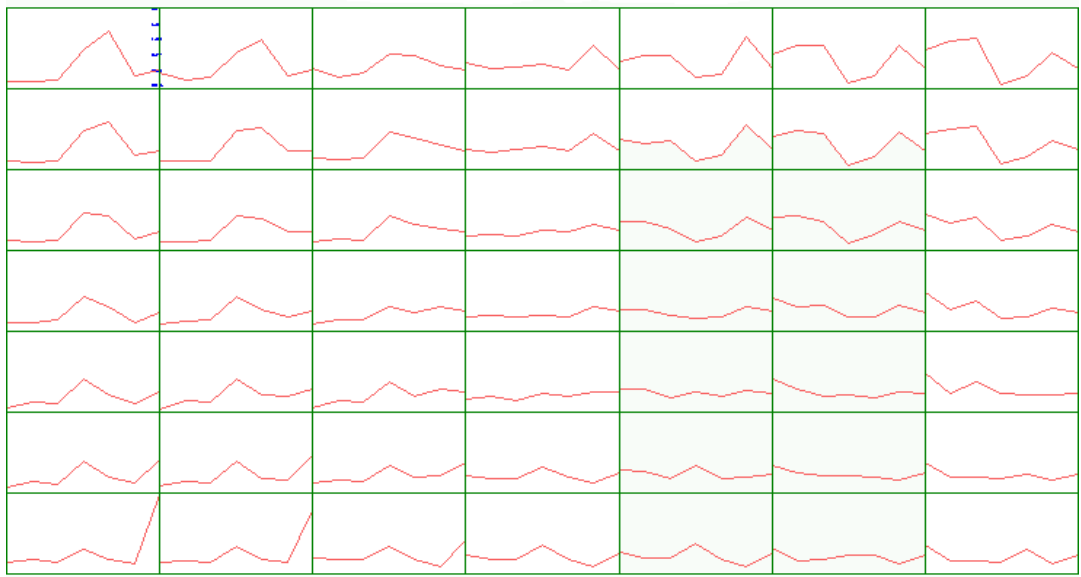

*Figure 1 : Répartition des vecteurs codes après convergence de l'algorithme (1000 itérations).*

*Figure 2 : Classement des 166 pays ayant au maximum une donnée manquante.*

La figure 3 montre le classement des 16 pays restant, classés comme des individus supplémentaires. On voit que la logique est respectée. Monaco est classé avec les pays riches, et la Guinée avec les pays très pauvres, ... Il peut paraître étrange qu'Andorre soit placé à l'opposé des pays riches, mais pour ce pays, on ne dispose pas de la variable PNBH, ce qui l'éloigne de l'ensemble des pays riches.

*Figure 3 : Les pays ayant 2 valeurs manquantes ou plus sont classés comme données supplémentaires.*

Si on représente le niveau (de 1 à 6) de la variable NIVIDH sur la carte, on voit (figure 4) que les pays sont effectivement classés selon la logique de cet indice ; La classification n'est pas parfaite, ce qui est normal, car cet indice de développement prend en compte d'autres variables plus liées à la qualité de la vie (comme le nombre de médecins, de cinémas, la sécurité, etc.).

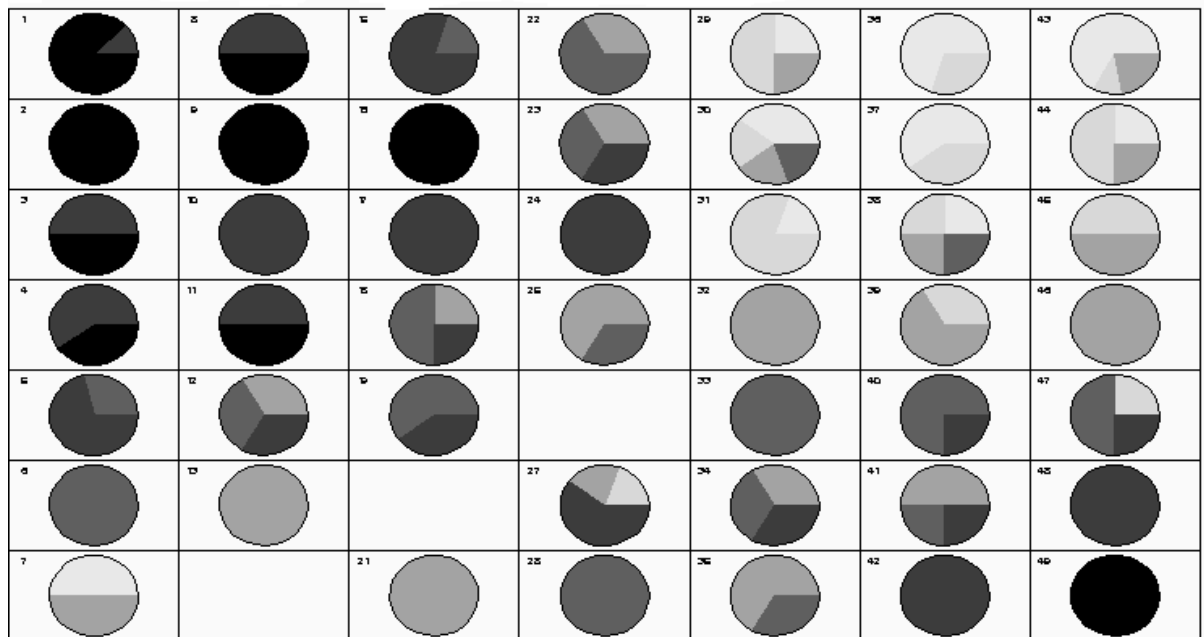

*Figure 4 : Proportion des modalités de la variable NIVIDH classée en 6 niveaux selon la classe. L'échelle va de 6 (gris très foncé) à 1 (gris très clair).*

On remarque que l'introduction de 52 pays incomplets ne produit aucune incohérence. En ce qui concerne les pays considérés en données supplémentaires, si leur positionnement sur la carte est logique (figure 3), il est difficile de les inclure dans le graphe ci-dessus, parce qu'on ne dispose pas pour eux de l'indice de développement humain.

## 6. ETUDE DU MARCHE IMMOBILIER FRANCILIEN

Il s'agit d'une étude pour le compte de la direction de l'habitat et de la ville de la Direction Régionale de l'Equipement d'Ile-de-France (DHV/DREIF), réalisée en 1993 par le laboratoire METIS de Paris 1, en collaboration avec le SAMOS.

On dispose pour 205 communes d'Ile-de-France étudiées en 1988, de données immobilières (loyers et prix des logements, anciens ou neufs, collectifs ou individuels, standard ou luxe, loyers et prix des bureaux, anciens ou neufs). Structurellement, certaines données sont absentes, par exemple le marché des bureaux peut être inexistant dans certaines communes.

Il s'agit de données où les données manquantes sont donc manquantes de manière structurelle, et où le nombre de communes se réduit drastiquement si l'on supprime celles qui sont incomplètes, on ne pourrait en considérer que 5 sur les 205 !

On représente dans la figure 5, la liste des communes (avec ou sans données manquantes) classées sur une carte de Kohonen. Remarquons que sur l'ensemble de la table des données, il y a environ 63% de valeurs manquantes.

| | 94_ST MAUR | 94_JOINVILL | 94_BRY S/MA<br>95_PONTOISE<br>77_DAMMARIE<br>77_OZOIR LA | 91 ATHIS MO<br>91 BRUNO/<br>9 MORSANG<br>91 VIRY CHA | 91 BRETIGNY<br>91 DRAVEIL<br>91 LES ULIS<br>91 RIS ORAN<br>91 ST MICHE<br>77 PROVINS<br>91 ROISSY E | 93 AUBER<br>93 BEZON<br>95 VILLI<br>91 CORBE<br>91 GIF S<br>91 MASSY<br>91 ORSAY<br>91 PALAI | 91 VIGNE<br>77 CHAMP<br>77 CHELL<br>77 LAGNY<br>77 LESIG |
|---|---|---|---|---|---|---|---|
| | | | 78, 94 | 94, 94, 95, 95 | 94, 94, 95, 91, 77 | 95, 95, 95, 91, 91 | 93, 95, 91, 77 |
| 75_PARIS 18<br>75_PARIS 19<br>*Boulogne* | | 92_ASNIERES<br>94_KREMLIN | 94_SUCY EN<br>95_DEUIL LA<br>93_PIERREFI<br>78_VILLE NE<br>91_SAVIGNY/O | 94_LE PLEST<br>95_L'ISLE L<br>95_ERMONT<br>95_TAVERNY<br>91_EPINAY=O<br>91_JUVISY S<br>77_MELUN | 94_CHEVILLY<br>95_L'ISLE L<br>78_LES MURE<br>91_EVRY<br>91_YERRES | 95_CERGY<br>95_FRANC<br>95_MONTI<br>95_SANNO<br>95_ST BR<br>91_BURES<br>91_VERRI<br>77_COULO<br>77_LE ME | 77_SAVIG<br>91, 91, 93, 94 |
| | | | 94, 94, 95, 77 | 91, 95, 95, 95 | 95 | | |
| 75_PARIS 10<br>75_PARIS 12<br>75_PARIS 13 | 75_PARIS 20 | 92_BOURG LA<br>92_COURBEVO<br>92_FONTENAY<br>92_LE PLESR<br>92_PUTEAUX<br>92_SURESNES | 92_COLOMBES | 94_VITRY S/<br>93_BAGNOLET<br>93_NOISY LG<br>95_ENGHIEN<br>78_POISSY<br>78_RAMBOUIL<br>91_STE GENE | 78_MANTES | 95_ARGENTEU<br>91_HEREBLAV<br>95_MONTMORE<br>78_CONFLANS<br>91_ETAMPES<br>91_COMBES LA<br>77_NEMOURS<br>77_PONTAULT | 93, 93 |
| | | | 93 | 94, 94 | | 94, 94, 93, 93, 95 | |
| 75_PARIS 2<br>75_PARIS 3<br>75_PARIS 4 | 75_PARIS 11<br>92_LEVALLOI | 94_CHARENTO | 94_LE PERRE<br>78_VERSAILL | 92_CLICHY<br>94_CRETEIL<br>78_HOUILLES | 92_GENNEVIL<br>92_CHOISY L<br>93_BOBIGNY<br>94_L'HAY LE | 94_CHENNEVI<br>77_AVON<br>77_FONTAINE<br>77_MEAUX | |
| | | | | | 94, 94 | 94, 93, 93 | |
| 75_PARIS 5<br>75_PARIS 14<br>75_PARIS 17<br>92_NEUILLY | | 92_SCEAUX | 92_GARCHES<br>92_MEUDON<br>92_ST CLOUD<br>94_ST MAURI | 92_NANTERRE<br>78_SAINT GE | 94_CACHAN<br>93_MONTREU | 78_MAISON-L | |
| | | | | | 92 | | |
| 75_PARIS 16<br>75_PARIS 9<br>75_PARIS 15 | 75_PARIS 1 | 94_ST MANDE | 92_VANVES<br>94_NOGENT S | 92_CLAMART<br>92_RUEIL MA<br>94_FONTENBO | | 92_ANTONY<br>92_CHATILLO<br>94_MAISON-A<br>93_PANTIN | |
| 75_PARIS 6<br>75_PARIS 7<br>75_PARIS 8 | | 92_VILLE D' | 92_CHAVILLE<br>92_ISSY LES<br>92_SEVRES<br>94_VINCENNE | | 92_BOIS COL<br>92_MALAKOFF<br>92_MONTROUG | 92_LA GAREN | |
| | | | 92 | 92 | 92 | 94 | |

*Figure 5 : Classement des communes, on voit clairement à gauche Paris, puis la petite couronne, et ensuite la grande couronne. Les numéros des départements représentent les 55 communes ayant plus de 12 valeurs manquantes, qui ont été classées après coup, comme données supplémentaires.*

On voit dans cet exemple, pratiquement impossible à traiter avec un logiciel classique, que l'algorithme de Kohonen permet de classer malgré tout des données extrêmement « trouées » sans introduire d'erreur grossière.

Bien sûr, ces bons résultats s'expliquent par le fait que les 15 variables mesurées sont bien corrélées, et donc les valeurs présentes contiennent de l'information sur les valeurs manquantes. La table suivante contient la matrice de corrélation des 15 variables observées.

|  1    | 0.77 | 0.77 | 0.68 | 0.86 | 0.73 | 0.67 | 0.91 | 0.74 | 0.71 | 0.67 | 0.86 | 0.71 | 0.85 | 0.78 |
|---|---|---|---|---|---|---|---|---|---|---|---|---|---|---|
| 0.77 |  1    | 0.97 | 0.83 | 0.85 | 0.85 | 0.86 | 0.90 | 0.85 | 0.83 | 0.86 | 0.87 | 0.83 | 0.93 | 0.99 |
| 0.77 | 0.97 |  1    | 0.81 | 0.83 | 0.84 | 0.88 | 0.86 | 0.83 | 0.83 | 0.87 | 0.83 | 0.83 | 0.92 | 0.99 |
| 0.68 | 0.83 | 0.81 |  1    | 0.99 | 0.71 | 0.95 | 0.99 | 0.98 | 0.78 | 0.72 | 0.98 | 0.97 | 0.92 | 0.83 |
| 0.86 | 0.85 | 0.83 | 0.99 |  1    | 0.79 | 0.79 | 0.98 | 1.00 | 0.75 | 0.70 | 1.00 | 0.97 | 0.92 | 0.86 |
| 0.73 | 0.85 | 0.84 | 0.71 | 0.79 |  1    | 0.92 | 0.82 | 0.72 | 0.96 | 0.99 | 0.81 | 0.66 | 0.81 | 0.86 |
| 0.67 | 0.86 | 0.88 | 0.95 | 0.79 | 0.92 |  1    | 0.92 | 0.79 | 0.99 | 0.91 | 0.80 | 0.97 | 0.72 | 0.88 |
| 0.91 | 0.90 | 0.86 | 0.99 | 0.98 | 0.82 | 0.92 |  1    | 0.96 | 0.87 | 0.83 | 0.98 | 0.96 | 0.97 | 0.89 |
| 0.74 | 0.85 | 0.83 | 0.98 | 1.00 | 0.72 | 0.79 | 0.96 |  1    | 0.72 | 0.65 | 0.98 | 0.99 | 0.91 | 0.85 |
| 0.71 | 0.83 | 0.83 | 0.78 | 0.75 | 0.96 | 0.99 | 0.87 | 0.72 |  1    | 0.91 | 0.76 | 0.79 | 0.75 | 0.84 |
| 0.67 | 0.86 | 0.87 | 0.72 | 0.70 | 0.99 | 0.91 | 0.83 | 0.65 | 0.91 |  1    | 0.71 | 0.63 | 0.79 | 0.86 |
| 0.86 | 0.87 | 0.83 | 0.98 | 1.00 | 0.81 | 0.80 | 0.98 | 0.98 | 0.76 | 0.71 |  1    | 0.96 | 0.92 | 0.87 |
| 0.71 | 0.83 | 0.83 | 0.97 | 0.97 | 0.66 | 0.97 | 0.96 | 0.99 | 0.79 | 0.63 | 0.96 |  1    | 0.92 | 0.84 |
| 0.85 | 0.93 | 0.92 | 0.92 | 0.92 | 0.81 | 0.72 | 0.97 | 0.91 | 0.75 | 0.79 | 0.92 | 0.92 |  1    | 0.93 |
| 0.78 | 0.99 | 0.99 | 0.83 | 0.86 | 0.86 | 0.88 | 0.89 | 0.85 | 0.84 | 0.86 | 0.87 | 0.84 | 0.93 |  1    |

On voit que les coefficients sont très élevés, 76 (sur 105) sont supérieurs à 0.8, aucun n'est inférieur à 0.65.

# 7. STRUCTURE DE DEPENSES DE L'ETAT, DE 1872 A 1971

Pendant 24 années, séparées en 3 catégories (avant la guerre 14-18, entre les deux guerres, après la seconde guerre), on a mesuré 11 variables représentant les dépenses de l'Etat dans différents secteurs : Pouvoirs publics, Agriculture, Commerce et industrie, Transports, Logement et aménagement du territoire, Education et culture, Action sociale, Anciens combattants, Défense, Dette, Divers.

Il s'agit donc d'un petit exemple, avec 24 observations de dimensions 11, sans valeurs manquantes.

Une analyse en composantes principales toute simple permet une excellente représentation même en deux dimensions (64 % de variance expliquée). Voir figure 6, la représentation des variables et figure 7, la représentation des années.

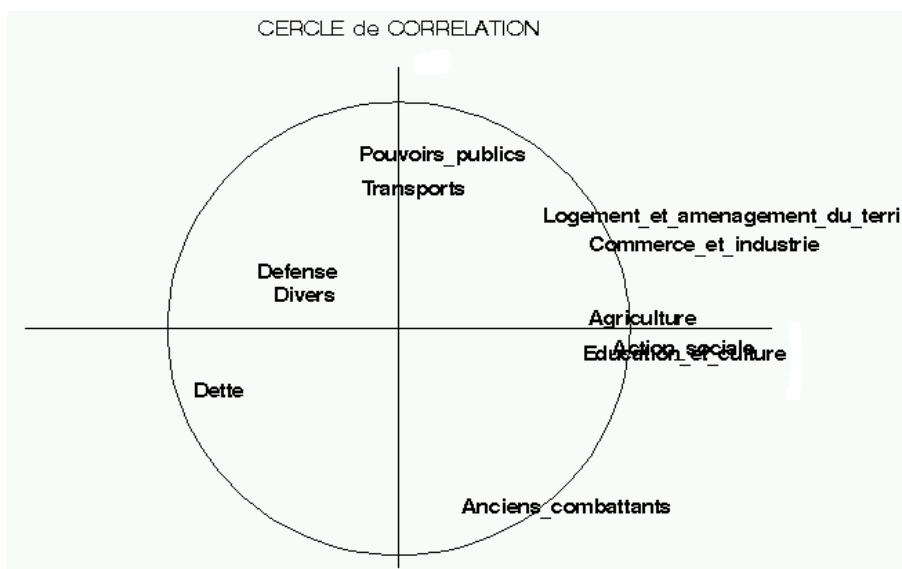

*Figure 6 : Représentation des variables sur le premier plan principal.*

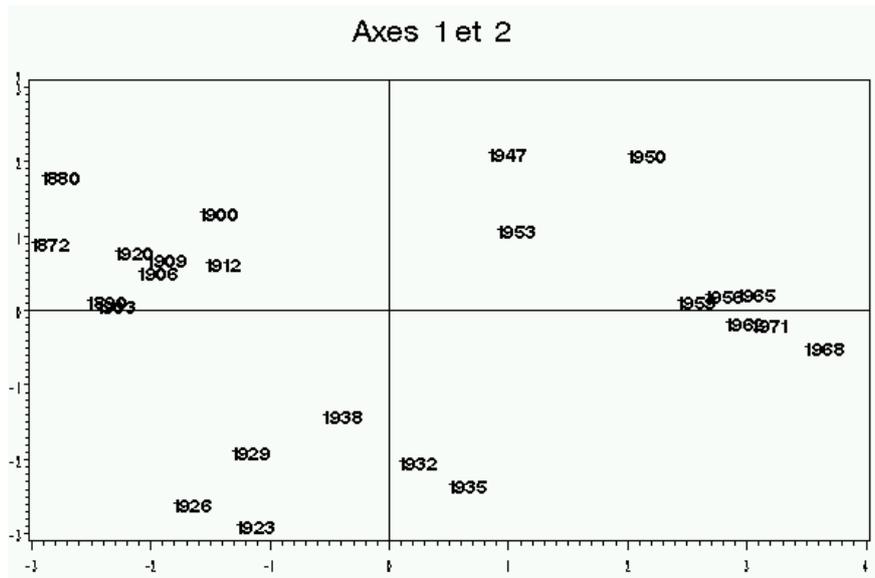

*Figure 7 : Représentation des années sur le premier plan principal. On distingue parfaitement les trois groupes.*

On remarque que les années se répartissent en trois groupes, qui correspondent aux trois périodes clairement définies (avant la première guerre mondiale, entre les deux guerres, après la seconde guerre mondiale). Seule l'année 1920, première année où il apparaît un poste de dépenses consacré aux anciens combattants est placée avec le premier groupe, alors qu'elle appartient au second.

Sur la figure 8, les années sont représentées sur une carte de Kohonen de taille 3, 3 et on observe les mêmes regroupements. On a regroupé les années en trois classes à l'aide d'une classification hiérarchique des vecteurs codes.

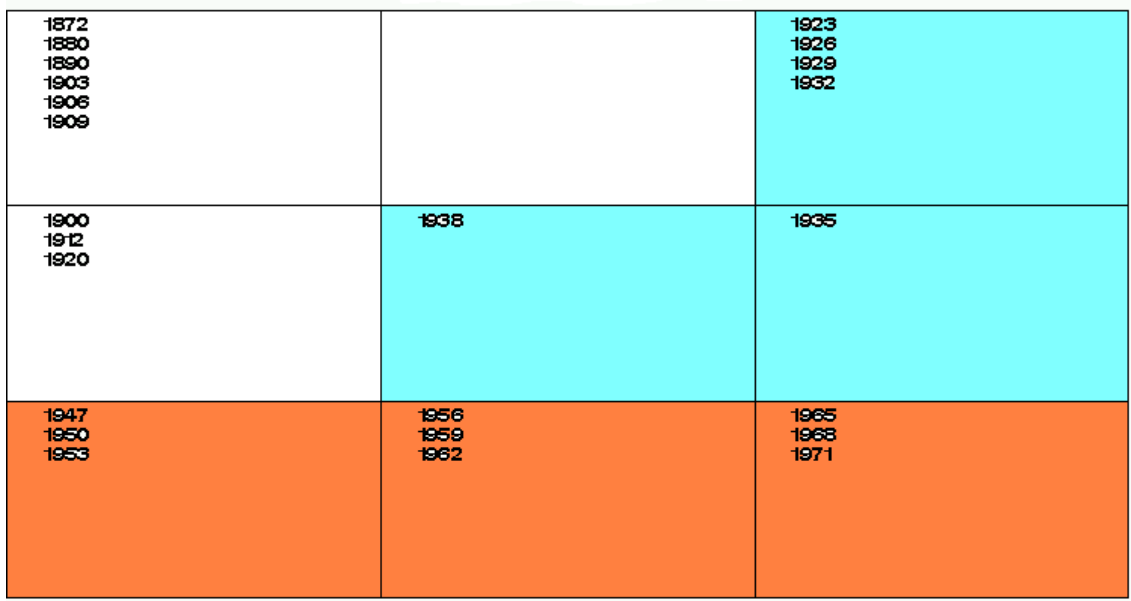

*Figure 8 : Carte de Kohonen des 24 années, 1920 est classé avec le premier groupe.*

Dans la figure suivante, on représente les cartes de Kohonen obtenues après avoir artificiellement supprimées un certain nombre de valeurs. Au lieu d'écrire la date précise, on note 1, 2 ou 3, selon la période. On supprime de 1 à 8 valeurs par année (sur 11 au total) aléatoirement. Les cartes obtenues, avec le regroupement en trois classes par la méthode de classification hiérarchique, sont présentées dans la figure 9.

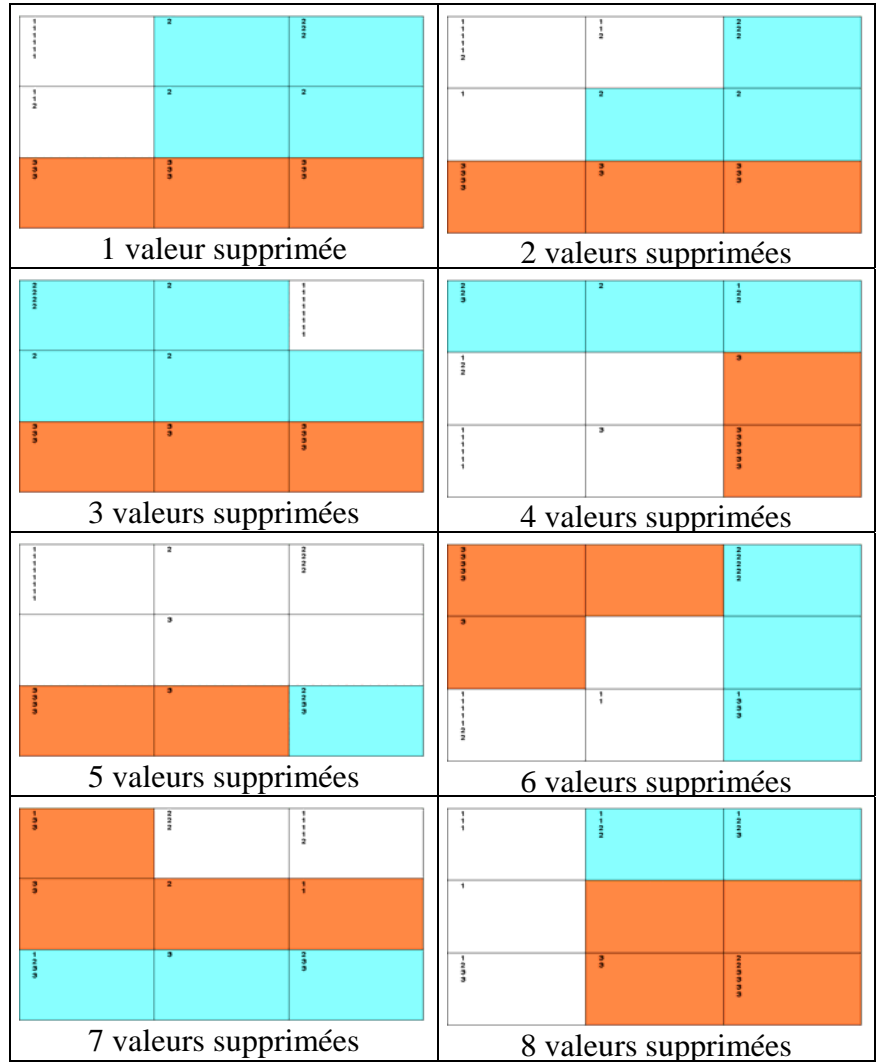

*Figure 9 : On voit que les super-classes restent homogènes tant qu'on ne supprime pas plus de 3 valeurs par année, soit 27% des valeurs. Ensuite les classes d'années se mélangent. Les années sont notées 1, 2, 3, suivant la période.*

On estime ensuite dans chaque cas les valeurs qui avaient été supprimées. Le tableau suivant montre dans chaque cas l'erreur quadratique moyenne. La figure 10 présente l'évolution de cette erreur en fonction du nombre de valeurs supprimées.

| Nb val | 1 | 2 | 3 | 4 | 5 | 6 | 7 | 8 |
|---|---|---|---|---|---|---|---|---|
| | 0.39 | 0.54 | 0.73 | 1.11 | 1.31 | 1.30 | 1.27 | 1.39 |

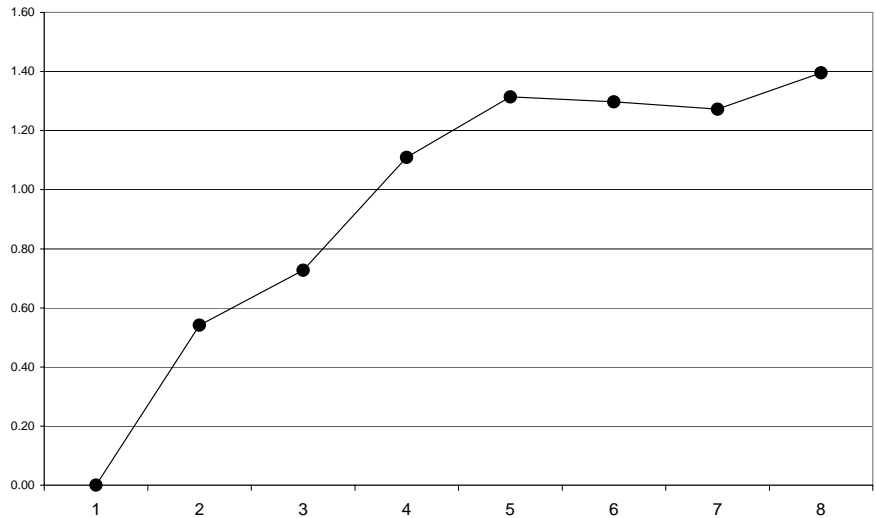

*Figure 10 : Erreur quadratique moyenne d'estimation en fonction du nombre de valeurs supprimées dans chaque année.*

On constate que l'erreur reste très faible lorsqu'on ne supprime pas plus 3 valeurs par année. Ensuite l'erreur plafonne, cela est dû au fait que le nombre de données (24) et de vecteurs codes (9) sont faibles tous les deux. Les composantes des vecteurs codes disponibles sont en nombre limité, et les estimations restent dans ce cas dans l'ensemble de ces composantes.

## 8. CONCLUSION

On a donc montré sur ces trois exemples comment il est possible et souhaitable d'utiliser des cartes de Kohonen quand les données disponibles présentent des données manquantes. Bien sûr les estimations et les classes obtenues seront d'autant plus pertinentes que les variables descriptives des données sont bien corrélées.

L'exemple 2 montre qu'il peut s'agir de la seule méthode possible lorsque les données sont extrêmement trouées. L'exemple 3 montre comment cette méthode permet d'estimer avec précision des données absentes. Les données ainsi complétées peuvent ensuite faire l'objet de tout traitement classique.

## 9. REFERENCES

Blayo, F., Demartines, P.(1991) : Data analysis : How to compare Kohonen neural networks to other techniques ? In *Proceedings of IWANN'91*, Ed. A.Prieto, Lecture Notes in Computer Science, Springer-Verlag, 469-476.

Bouroche et Saporta G. (1980) : *L'analyse des données*, Que-sais-je ? PUF, Paris.

Cottrell M., de Bodt E., Verleysen M. (2002), A statistical tool to assess the reliability of Self-Organizing Maps, *Neural Networks*, Vol 15, n°8-9, p967-978.

Cottrell M., Ibbou S., Letrémy P., Rousset P. (2003) : Cartes auto-organisées pour l'analyse exploratoire de données et la visualisation, à paraître dans le *Journal de la Société Française de Statistique.*

Ibbou S., (1998) : Classification, analyse des correspondances et méthodes neuronales, Thèse, Université Paris 1.

Gaubert, P., Ibbou, S., Tutin, C., (1996) :Segmented Real Estate Markets and Price Mechanisms : the Case of Paris, *International Journal of Urban and Regional Research,* 20, n°2, 270-298.